\documentclass[review]{elsarticle}

\usepackage{lineno,hyperref}
\modulolinenumbers[5]

\usepackage[version=3]{mhchem}
\usepackage{graphicx}
\usepackage{dcolumn}
\usepackage{bm}
\usepackage{threeparttable}
\usepackage{colordvi}
\usepackage{color}
\usepackage{booktabs}
\usepackage[english]{babel}
\usepackage{amsfonts}
\usepackage{graphicx}

\newcommand{\Eads}{E_{\rm ads}}

\newcommand{\Etot}{E_{\rm total}}
\newcommand{\Emol}{E_{\rm molecule}}
\newcommand{\Esur}{E_{\rm surface}}

\newcommand{\rhotot}{\rho_{\rm DBAQ/Au}}
\newcommand{\rhomol}{\rho_{\rm DBAQ}}
\newcommand{\rhosur}{\rho_{\rm Au}}

\begin{document}

\begin{frontmatter}

\title{Selective Adsorption of a Supramolecular Structure on Flat and Stepped Gold Surfaces}

\author[RPR]{Rengin Pek\"{o}z\corref{cor1}}
\ead{rengin.pekoz@atilim.edu.tr}

\author[UC,MPIP,IKER]{Davide Donadio\corref{cor2}}
\ead{ddonadio@ucdavis.edu}

\cortext[cor1]{Corresponding author}
\cortext[cor2]{Principal corresponding author}

\address[RPR]{Department of Electrical and Electronics Engineering, At{\i}l{\i}m University, 06836 Ankara, Turkey}
\address[UC]{Department of Chemistry, University of California Davis, One Shields Avenue, Davis, CA, 95616, U.S.A}
\address[MPIP]{Max Planck Institute for Polymer Research, Ackermannweg 10, Mainz 55128, Germany}
\address[IKER]{IKERBASQUE, Basque Foundation for Science, E-48011 Bilbao, Spain}

\begin{abstract}
Halogenated aromatic molecules assemble on surfaces forming both hydrogen and halogen bonds. Even though these systems have been intensively studied on flat metal surfaces, high-index vicinal surfaces remain challenging, as they may induce complex adsorbate structures. The adsorption of 2,6-dibromoanthraquinone (2,6-DBAQ) on flat and stepped gold surfaces is studied by means of van der Waals corrected density functional theory.~Equilibrium geometries and corresponding adsorption energies are systematically investigated for various different adsorption configurations.~It is shown that bridge sites and step edges are the preferred adsorption sites for single molecules on flat and stepped surfaces, respectively. The role of van der Waals interactions, halogen bonds and hydrogen bonds are explored for a monolayer coverage of 2,6-DBAQ molecules, revealing that molecular flexibility and intermolecular interactions stabilize two-dimensional networks on both flat and stepped surfaces. Our results provide a rationale for experimental observation of molecular carpeting on high-index vicinal surfaces of transition metals. 
\end{abstract}


\end{frontmatter}

\section{Introduction}
Supramolecular structures have been attracting great amount of attention due to their potential applications in materials technology, catalysis, medicine, and data storage and processing \cite{Rao2004,Elemans2009,Wehrspohn2011}. For this reason, the structural arrangements of large molecules on metal surfaces have been widely explored both experimentally and theoretically in the last two decades \cite{Rosei2003,Kim2012,Jang2014,Pham2014}. These studies have mostly concentrated on understanding the intermolecular interactions of well-ordered networks of organic molecules, such as hydrogen bonds, halogen bonds, dispersive interactions and dipole-dipole interactions. 

Halogen bonding, occurring between a donor halogen atom and an acceptor ligand, is a crucial tool in supramolecular chemistry \cite{Meazza2013,Gilday2015}. Even though hydrogen bonds are stronger than halogen bonds, one of the main advantages of halogen bonding is the flexibility it provides in tuning the binding properties of molecules on metals \cite{Liu2013,Pekoz2014}. Furthermore, it is accepted that halogen substitution, which cooperates between intermolecular and molecule-surface interactions, plays an important role in the formation of intermolecular networks on metal surfaces \cite{Jenny2012,Cui2013}. Among many functional molecules forming ordered and two-dimensional networks, dibromo- and dichloroanthraquinones molecules have recently attracted interest for the supramolecular structures that they form on flat gold surfaces \cite{Yoon2011b,Yoon2011,Noh2013}. 

Defects at surfaces, such as steps, kinks and vacancies, are also of great importance for the	 molecular self-assembling process.~Most of the studies have investigated the interaction of large molecules with atomically smooth flat surfaces \cite{Jang2014,Yoon2011b,Schnadt2010,Chung2013,Kepcija2013,Zha2014,Kawai2015,Liu2016}, whereas high-index metallic surfaces, occurring in realistic situations, have attracted little attention \cite{Schnadt2008,Waldmann2012,Kim2016}. It is known that it is easier to control the formation of molecular networks on perfect flat surfaces \cite{Yokoyama2001,Keeling2003,Schiffrin2007}, whereas on high-index vicinal surfaces it is more challenging to form uniform supramolecular structures.~Few experimental studies have targeted the formation of one-dimensional (1D) \cite{Schnadt2010,Schnadt2008} and two-dimensional (2D) supramolecular assemblies \cite{Kim2016} on stepped surfaces, which have raised the necessity to understand the interactions taking place between organic molecules and stepped surfaces. Intermolecular interactions among dibromoanthraquinone (DBAQ) molecules have been theoretically studied on flat Au(111) surface \cite{Yoon2011b}, however, up to our knowledge, the most favorable adsorption site and geometry of DBAQ molecule on flat and stepped gold surfaces, as well as the interaction of DBAQ with the substrate have not been studied.

In this work, we explore the interaction of a large organic molecule, 2,6-dibromoanthraquinone (2,6-DBAQ), with flat (111) and stepped gold surfaces, specifically (322) and (443), by means of density functional theory calculations with nonlocal van der Waals (vdW) correlation functional. The main purpose of this work is to elucidate the role of intermolecular interactions, such as hydrogen and halogen bonding, and vdW interactions with different surfaces, in the formation of self-assembled structures of DBAQ molecules on gold surfaces. Adsorption sites and energies as a function of coverage and electronic structures are studied, allowing a detailed understanding of the interaction of DBAQ molecule with Au surfaces.~We show that the inclusion of vdW forces is crucial to describe the adsorption of DBAQ molecule on gold surfaces.~Due to the higher reactivity of stepped surfaces, the adsorption is more favorable on step edges than on terrace sites. Furthermore, the energy gain obtained due to the intermolecular halogen and hydrogen bonds for monolayer structures is partly compensated by the weaker interaction of DBAQ molecules with the gold surfaces.~As a result, these interactions cooperatively stabilize 2D networks not only on flat (111) but also on stepped (322) gold surfaces. 

After a brief summary of the computational details used in the calculations, in section 3.1 the adsorption of a single 2,6-DBAQ on Au(111) surface with different adsorption sites, energies and heights as a function of coverage will be discussed. In section 3.2, adsorption properties on stepped Au(322) will be investigated.~In section 3.3, the details of the supramolecular carpeting on both surface types will be explored. The electronic properties, including the charge density and partial density of states, of the systems studied will be analyzed and discussed in section 3.4.  \\

\section{Computational Methods}
The adsorption of 2,6-dibromoanthraquinone (DBAQ) on flat and stepped surfaces is investigated by density functional theory (DFT). The generalized gradient approximation (GGA) with Perdew-Burke-Ernzerhof (PBE) \cite{Perdew1996,Perdew1998} exchange-correlation energy functional is used. Since the long-range van der Waals (vdW) interactions play an important role, especially for planar configurations, on the adsorption energy and molecule-surface geometry, their particular role is investigated by means of the approaches proposed by Dion et al. \cite{Dion2004}. Even though significant developments to treat dispersion forces have been carried out since, we have verified that this first-generation vdW-DF functional yields a very reliable description of the adsorption energy and geometry of aromatic compounds at metal surfaces \cite{Pekoz2014,Klimes2011,GLi2012,Pekoz:2012tt,Berland2014}. The electronic wave functions are expanded in a plane-wave basis set with an energy cutoff set to 400 eV, and the core electrons are described by the frozen-core all-electron projector-augmented wave (PAW) potentials \cite{Blochl1994,Kresse1999a}. All the calculations were performed using the VASP code \cite{Kresse1996b,Kresse1996c}.  

The lattice constant of Au with vdW-DF exchange-correlation functional is found to be 4.21 \AA, which is close to the experimental value of 4.06 \AA~\cite{Auexp} and previously reported theoretical data \cite{Khein1995,Carrasco2014}.~The (111) and (322) and (443) surfaces are considered as model flat and stepped surfaces, respectively, even though gold may exhibit more complex reconstructions at steps \cite{Gaspari:2010dz}. In particular, we consider Au(443) surface to compare the adsorption of DBAQ at a proper terrace to a flat (111) surface. Both flat and stepped surfaces are modeled by slabs four atomic layers thick, with the two bottom layers fixed during the simulations to mimic the bulk. The rest of the system is fully optimized with a convergence criterion of 0.1 meV for the energy and 10 meV/\AA\ for the forces. 

We tested the convergence of adsorption energies and structures with respect to the size of the simulation cell for the (111) surface, considering three supercells, the dimensions of which are reported in Table \ref{T:celldim}. For a given adsorption site, increasing the surface supercell area from 1.39 nm${^2}$ to 3.72 nm${^2}$ increases the binding energy by 0.06 eV, due to the interaction of DBAQ with its periodic images.~The height of the simulation cell is 30 \AA, which assures that the interaction among periodic replicas of the slab is negligible. Brillouin zone integration was performed using the Monkhorst-Pack scheme \cite{MPkpt} with 5$\times$3$\times$1 and 4$\times$2$\times$1 meshes for the (111) surface with (3$\times$3$\sqrt{3}$) supercell and (322) stepped surface with (6$\times$1) supercell, respectively. When different supercells are used, the size of the k-point meshes is accordingly scaled.    

The adsorption energy, $\Eads$, is defined as
\begin{equation}
\Eads = \Etot - \Esur - \Emol
\end{equation}
where $\Etot$, $\Esur$, and $\Emol$ are the total energies of the adsorbed system, the clean Au surface, and the isolated molecule, respectively. Consequently, a negative adsorption energy tells that the adsorbed system is energetically more favorable compared to the isolated state.   

DBAQ is a planar molecule and is expected to interact weakly with the gold surface through the $p$-electrons of the aromatic rings. To test the effect of dispersion forces, the adsorption energy of a single DBAQ molecule on Au(111) surface with B-30 adsorption site is calculated using the conventional PBE functional, which does not include the dispersion corrections. The adsorption energy is found to be -0.10 eV and the adsorption height to be 3.85 \AA~with PBE functional, while those calculated with vdW-DF functional are -2.08 eV and 3.38 \AA, respectively. The weak physisorption found with PBE functional shows that the adsorption of DBAQ on flat gold surfaces should be indeed governed by vdW interactions.~Thus, the calculations presented in this work are performed with vdW-DF functional, unless otherwise noted.

\begin{table}[h]
\caption{ Supercell sizes and number of Au atoms per supercell.}
\label{T:celldim}
\begin{tabular}{cccrr} \hline
 Surface &  $\#$ of atoms    & cell size      &  a [\AA]      &   b [\AA]   \\
 \hline
 Au(111) &        72                & (3$\times$3$\sqrt{3}$) &  8.97        &  15.54      \\
               &        96               & (6$\times$2$\sqrt{3}$) &   17.94      &   10.36   \\   
              &        192              & (6$\times$4$\sqrt{3}$) &  17.94      &   20.72     \\
Au(322) &         120             & (6$\times$1) &  17.94     &  12.32     \\
Au(443) &         180             & (6$\times$1) & 17.94       &   19.15    \\
\hline
\end{tabular}
\end{table}

\begin{figure}[h]
\centering
\includegraphics[width=65mm]{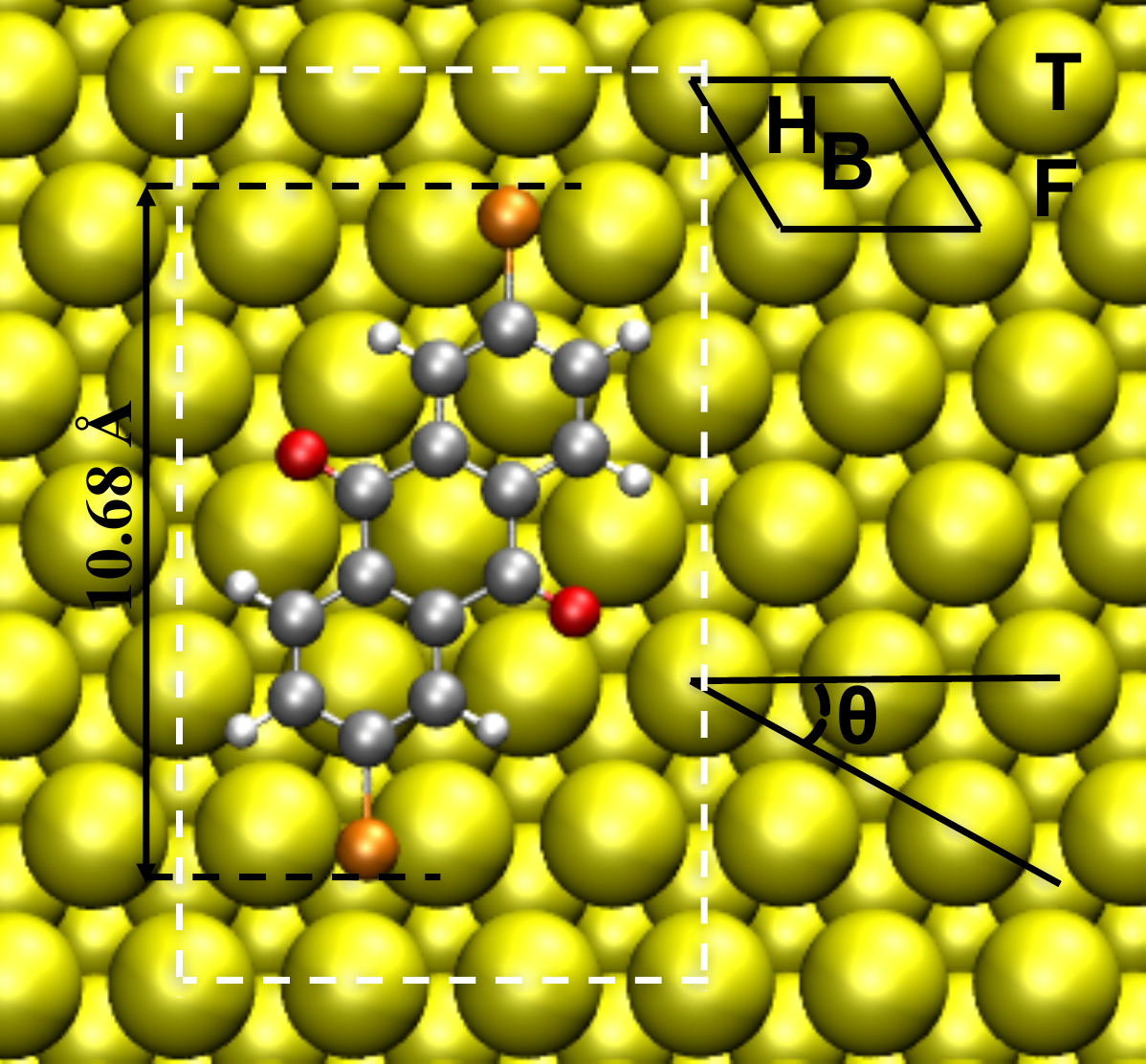}
\caption{Au(111) surface showing the high-symmetry adsorption geometries of 2,6-DBAQ molecule with the position of the center of C-ring is located on the top (T), bridge (B), fcc (F) and hcp (H) hollow sites. O-O position with respect to the surface is given by the azimuthal angle ($\theta$) of the molecule from the direction of Au rows. The (3$\times$3$\sqrt{3}$) supercell is drawn with white dashed line. Orange, red, white, gray and yellow spheres represent Br, O, H, C and Au atoms, respectively.}
\label{fig:dbaq+Au111_sites}
\end{figure}

\section{Results and Discussion}
The chemical structure of 2,6-dibromoanthraquinone molecule (DBAQ) is shown in Fig. \ref{fig:dbaq+Au111_sites}.~The vertical length between two Br atoms and the horizontal width between the oxygens of the molecule are 10.68 and 5.43 \AA, respectively.    

\subsection{DBAQ on Au(111)}
We perform the search for the most favorable adsorption site on Au(111) using an orthorhombic supercell with a 17.94 \AA~$\times$ 20.72 \AA~ section, which corresponds to a (6$\times$4$\sqrt{3}$) replica of the unitary surface cell. The adsorption of a single DBAQ molecule on flat Au(111) surface is explored for flat-lying orientation which has been suggested by the experimental studies \cite{Yoon2011b,Yoon2011}. The center of the molecule is positioned on four different adsorption sites (top, bridge, fcc and hcp) with two different orientations (0 and 30$^\circ$) with respect to the surface cell (see Fig. \ref{fig:dbaq+Au111_sites}). The O-O vector is chosen as the axis of the molecule. Among the eight different adsorption sites for a (6$\times$4$\sqrt{3}$) supercell, the most favorable sites are T-0 and B-30 with almost iso-energetic adsorption energies of -2.02 eV. The least energetic site is T-30 site with 0.18 eV adsorption energy difference with respect to the T-0 site. The other sites have very close adsorption energies to each other and E$_{ads}$ of hcp-30 site is only 0.03 eV lower than that of T-0 site. Such small energy differences among adsorption configurations are within the accuracy of our DFT calculations.~Therefore, our results can not be conclusive regarding the most stable adsorption site, but they show that the potential energy surface of DBAQ on Au(111) exhibits a few minima with very small energy differences. Hence, lateral and rotational diffusion of DBAQ on Au(111) are possible even at very low temperatures. DBAQ keeps its flat geometry upon adsorption for any adsorption site. The equilibrium height of each configuration, which is always larger than 3.3 \AA\ (see Table \ref{T:DBAQads+Au111}) indicates that DBAQ is physisorbed. 

\begin{table}[th]
\caption{Adsorption sites, angles ($\theta$), energies ($\Eads$), and average distances between the C-rings and the surface atoms (z$_C$), nearest atomic distances of oxygens (d$_O$) and bromines (d$_{Br}$) with the surface Au atoms of 2,6-DBAQ on Au(111) surfaces with (6$\times$4$\sqrt{3}$) supercells. The optimized configurations for each site is presented in Fig. S1, Supplementary materials.}
\label{T:DBAQads+Au111}
\begin{tabular}{crcccc} \hline
 Site  &  $\theta$ [$^{\circ}$]   &  $\Eads$ [eV]   & z$_C$ [\AA] & d$_O$  [\AA]  & d$_{Br}$  [\AA]   \\
 \hline
T       &      0                           &   {\bf -2.02}       &  3.34             & 3.15                &  3.61   \\
T       &       30                        &  -1.84               &  3.39             & 3.54/4.60        & 3.54  \\
hcp   &     0                            &  -1.95               &  3.40             & 3.65                & 3.65/3.50  \\
hcp   &    30                           &  -1.99               &  3.38             & 3.26/3.49        & 3.69/3.47   \\
fcc    &     0                            &  -1.95               &  3.41             & 3.66                & 3.51/3.66  \\
fcc    &   30                            &  -1.98               &  3.37             & 3.49/3.27        & 3.46/3.69  \\
B      &        0                         &  -1.94               &  3.42             & 3.60                & 3.38  \\
B      &      30                         &  {\bf -2.01}        &  3.37             & 3.55                & 3.39  \\  
\hline
\end{tabular}
\end{table}

\begin{figure}[h]
\centering
\includegraphics[width=120mm]{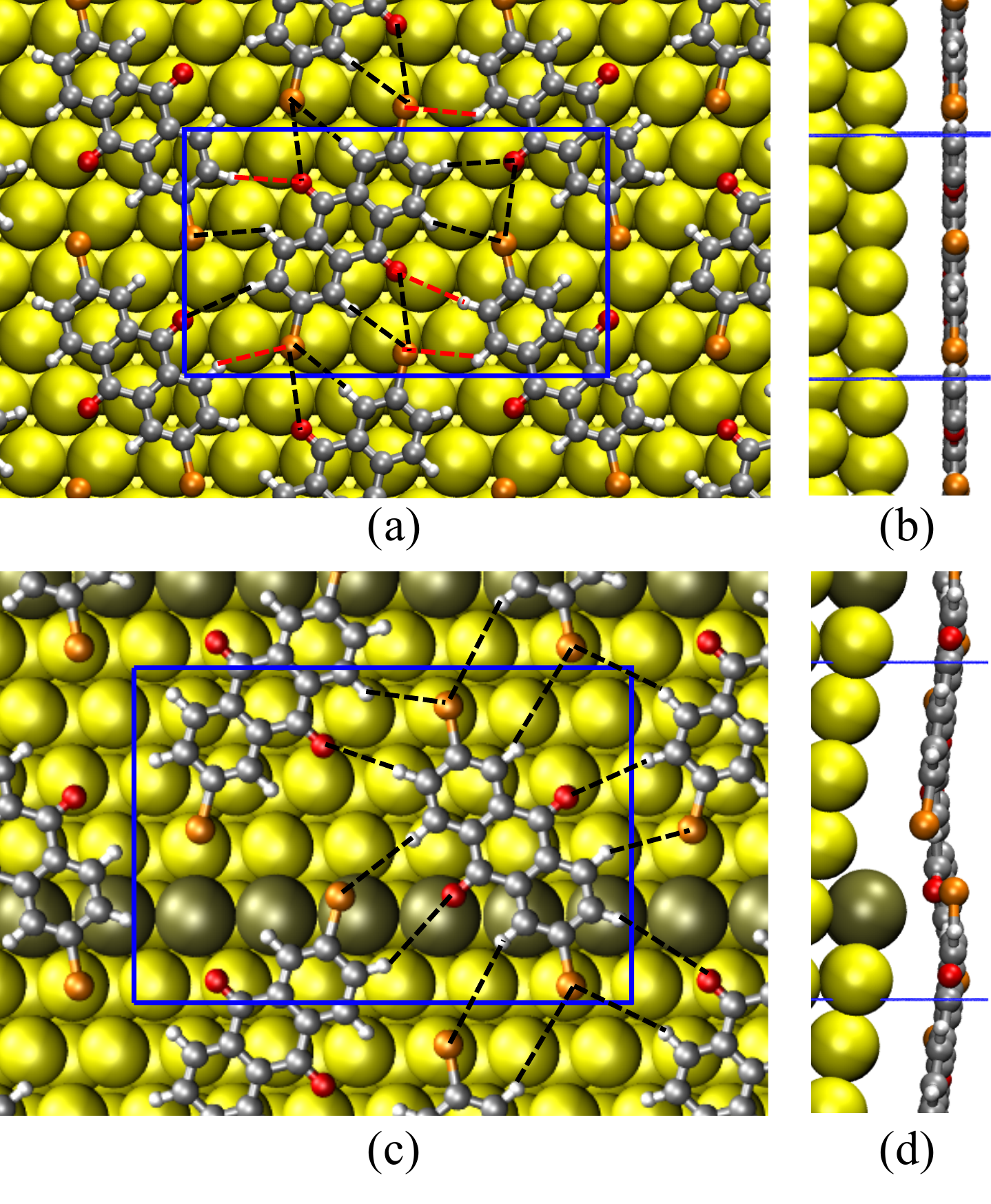}
\caption{Optimized geometries of two DBAQ adsorbed on (a-b) Au(111) with 6$\times$2 supercell and (c-d) Au(322) with 6$\times$1 supercell. The supercells are drawn with blue solid lines. The lattice constants are tabulated in Table \ref{T:celldim}. A zoom-in scheme of the configuration with the possible intermolecular bond lengths is presented in Fig. S5 and S6. }
\label{F:2dbaq+Au111}
\end{figure}

In order to understand the effect of bonding patterns in DBAQ self assembled monolayers, different cell sizes are considered for adsorption of a single DBAQ molecule with B-30 site as shown in Fig. S2. In the (6$\times$4$\sqrt{3}$) supercell, which is large enough to host a single DBAQ molecule without interactions with the neighboring images, the adsorption energy of DBAQ is -2.02 eV. 
In the (3$\times$3$\sqrt{3}$) supercell, DBAQ molecule forms a weak hydrogen bond with its neighboring image and the resulting adsorption energy is stronger by 0.06 eV than that in the (6$\times$4$\sqrt{3}$) supercell. Arranging DBAQ in a (6$\times$2$\sqrt{3}$) supercell allows a pattern, in which neighboring molecules interact through two Br$\cdot$$\cdot$$\cdot$H and two Br$\cdot$$\cdot$$\cdot$O bonds with an adsorption energy equal to -2.20 eV/molecule. This result shows that intermolecular Br$\cdot$$\cdot$$\cdot$H and Br$\cdot$$\cdot$$\cdot$O bonds may contribute significantly to the stability of DBAQ assemblies. The detailed bonding schemes are shown in Fig. \ref{F:2dbaq+Au111}(a). 

\subsection{DBAQ on Au(322)} 
\begin{figure}[h]
\centering
\includegraphics[width=85mm]{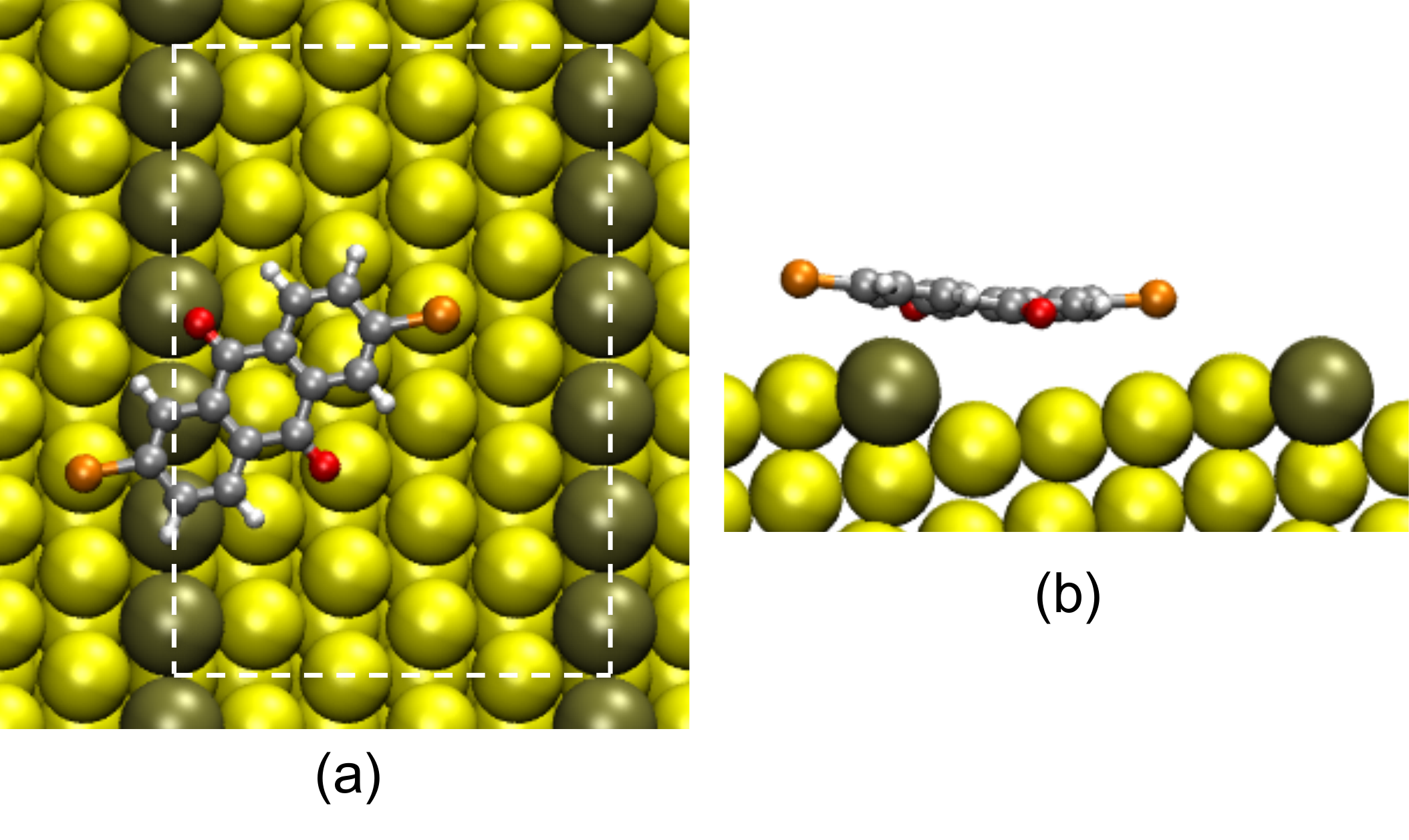}
\caption{The most favored adsorption configuration of DBAQ on Au(322) with (a) top view and (b) side view.~The step-edge gold atoms are displayed with larger spheres and darker color. The 6$\times$1 supercells are drawn with white dashed lines.~The calculated adsorption energy is -1.97 eV.}
\label{fig:dbaq+Au322_config}
\end{figure}

Here we consider an isolated 2,6-DBAQ molecule adsorbed on Au(322). Due to the lower symmetry of the (322) surface, there are many possible inequivalent adsorption sites. To understand the details of the molecular arrangement, seven different adsorption configurations (see Fig. S3) with the molecule on terrace and step sites with varying adsorption angles have been considered for a 6$\times$1 supercell. The most energetically favored configuration with the adsorption energy -1.97 eV is shown in Fig. \ref{fig:dbaq+Au322_config}. The molecule is tilted from step edge to lower terrace. The O-O molecular axis is at an angle of 30$^{\circ}$ with the step edge, so that one O atom is bonded to a gold atom at the step and the other points to the terrace. 
The nearest distance between O and Au atom of the step edge is 2.63 \AA. The side view of DBAQ adsorbed at the step edge clearly shows that the slanting adsorption of DBAQ stems from both the interaction of oxygen with an undercoordinated step atom and the surface{-}molecule interaction localized near the step edges, which is revealed by the electronic structure analysis and is explained more in detail in section 3.4. Decreasing the super cell size to 3$\times$1 resulted in an increase of adsorption energy by 0.17 eV, which is due to the introduction of O$\cdot$$\cdot$$\cdot$H and Br$\cdot$$\cdot$$\cdot$H bonds with the neighboring images. The configuration in Fig. \ref{fig:dbaq+Au322_config} is assumed as the starting geometry to study the surpramolecular carpeting of DBAQ molecules on Au(322), which is discussed in the following section.

The terraces of (322) and (443) stepped surfaces hold a (111) orientation and the steps are separated by four and seven atomic rows, respectively. Since the size of (443) surface with a (6$\times$1) supercell is comparable to that of the (6$\times$4$\sqrt{3}$) Au(111) supercell (see Table \ref{T:celldim}), adsorption properties at terrace sites of (443) surface might be expected to be similar to the flat one. The adsorption energy difference between (111) surface and (443) terrace  is only 0.06 eV and similar equilibrium distances (z$_{eq}$$\sim$3.65 \AA) are found. Adsorption energies and optimized geometries for the T30 adsorption site are presented in Fig. S4.  

Stepped surface are more reactive \cite{Smoluchowski1941}, as confirmed by the larger charge displacement ($\Delta Q$) upon adsorption and by the presence of shorter Au-O distances. Nevertheless the adsorption energies of single molecules, at the lowest coverages considered, are very similar for either flat Au(111) (-2.02 eV) or stepped Au(322) surface (-1.97 eV). The enhanced reactivity of the steps is compensated by distortions in the geometry of the molecule, which shifts the energy balance slightly in favor of flat surface or terrace adsorption.

\begin{table*}[h]
\caption{Surface types, adsorption sites, angles ($\theta$), energies ($\Eads$ in eV/DBAQ), equilibrium distances (z$_{eq}$ in \AA), nearest atomic distances of oxygens and bromines with the surface gold atoms (d in \AA), and charge displacement ($\Delta$Q in $e$) for DBAQ on Au surfaces calculated using vdW-DF functionals. }
\label{T:dbaq+Au}
\begin{tabular}{rlc ccccccc} 
\hline
                        &                             &                   &                      &\multicolumn{3}{c}{Terrace}                       & \multicolumn{2}{c}{Step edge} &                     \\
 \cmidrule(lr){5-7}\cmidrule(lr){8-9}                
System            &               Figure   & Supercell    & $\Eads$      &   d(O-Au)         &   d(Br-Au)   & z$_{eq}$  &       d(O-Au)    &  d(Br-Au)       &    $\Delta$Q    \\ 
\hline
DBAQ/Au(111) & S2(a) & (6$\times$4$\sqrt{3}$)  &      -2.02       &     3.26          &   3.34         &   3.39      &                       &       &   0.71    \\ 
DBAQ/Au(111) & \ref{fig:dbaq+Au111_sites}& (3$\times$3$\sqrt{3}$)  &      -2.08      &     3.35          &    3.40         &  3.43        &                      &                           &   0.49    \\
DBAQ/Au(111) & \ref{F:CDD}(a)   & (6$\times$2$\sqrt{3}$)  &      -2.20     &     3.36             &  3.42          & 3.41       &                       &                      &      0.72                \\
2DBAQ/Au(111) & \ref{F:2dbaq+Au111}(a)  & (6$\times$2$\sqrt{3}$)  &     -2.41      &     3.45       &   3.50         &  3.49       &             &                       &      0.98                \\  
DBAQ/Au(322) & \ref{fig:dbaq+Au322_config} & 6$\times$1  &   -1.97 &      3.55            &       3.52       &                  &    2.61    &   3.60       &      0.99                \\
2DBAQ/Au(322) & \ref{F:2dbaq+Au111}(c) &  6$\times$1 &   -2.13 &      3.72            &       3.54       &                 &    3.53     &   3.36      &      1.37                \\
\hline
\end{tabular}
\end{table*}

\subsection{Formation of a supramolecular carpet on Au(111) and Au(322)} 
Supramolecular carpeting of 2,6-DBAQ on Au(111) was studied experimentally by scanning tunneling microscopy, which showed a chevron-like pattern stabilized by halogen and hydrogen bonds \cite{Yoon2011}. In the same paper a detailed analysis of the intermolecular interactions for a suspended array of DBAQ molecules, i.e. without a substrate, was also performed using DFT calculations. In the light of this study, we have performed DFT calculations to examine the formation of chevron-like structure on Au(111) as well as Au(322) surfaces and confirm its stability considering the intermolecular and molecule-substrate interactions.

First we consider the formation of a supramolecular carpet on flat Au(111). To obtain a chevron-like structure on Au(111), two DBAQ molecules in alternating molecular rows on Au(111) surface have been examined. A (6$\times$2$\sqrt{3}$) Au(111) supercell is used, in agreement with the reported experimental and theoretical equilibrium lattice distances \cite{Yoon2011}. B-30 adsorption site, being the most energetic site on Au(111), is considered for the initial geometries of DBAQ molecules and the optimized configuration is shown in Fig. \ref{F:2dbaq+Au111}(a-b), yielding an adsorption energy of -2.41 eV/molecule. The intermolecular interactions in 2D network can be analyzed by considering four nearest neighboring DBAQ molecules. Each molecule forms with each neighbor two Br$\cdot$$\cdot$$\cdot$H and two O$\cdot$$\cdot$$\cdot$H bonds (see red dotted lines in Fig. \ref{F:2dbaq+Au111}(a)) as compared to the single DBAQ adsorption case in the same supercell. These additional bonds further stabilize the supramolecular structure by 0.21 and 0.40 eV per molecule, compared to single DBAQ adsorption in (6$\times$2$\sqrt{3}$) and (6$\times$4$\sqrt{3}$) supercells, respectively.~Upon adsorption on the B-30 sites, both DBAQ molecules maintain their flat geometry and the average distance between the surface atoms and the C-rings (z$_{eq}$) is 3.49 \AA. The increase of z$_{eq}$ by 0.10 \AA~along with the increase of $\Eads$ by 0.40 eV, compared to the  adsorption of an isolated molecule (in a (6$\times$4$\sqrt{3}$) supercell), suggests an interplay between the intermolecular interactions and the molecule-substrate interaction, so to maintain overall a constant bond order.

The intermolecular interactions can be understood by considering the bond lengths between neighboring molecules as presented in Fig. S5. O$\cdot$$\cdot$$\cdot$H bond lengths range from 2.88 to 3.21 \AA~being larger than the sum of the van der Waals radii for each bond (r$_{O-H}$=2.72 \AA), which suggest that hydrogen bonds formed between neighboring molecules are weak. Br$\cdot$$\cdot$$\cdot$O bond length is 3.49 \AA, while the range of Br$\cdot$$\cdot$$\cdot$H bond lengths are similar to those of O$\cdot$$\cdot$$\cdot$H bonds.~The sum of the vdW radii for Br-O is equal to r$_{Br-O}$=3.37 \AA, which is smaller than the calculated bond length, revealing that the intermolecular Br$\cdot$$\cdot$$\cdot$O bonds are weak.~On the other hand, r$_{Br-H}$=3.05 \AA, which is within the calculated range of Br$\cdot$$\cdot$$\cdot$H bond lengths (from 2.87 to 3.19 \AA), showing that the Br$\cdot$$\cdot$$\cdot$H bonds are relatively stronger than the other intermolecular bonds occurring in the self assembled monolayer.    

Then, using the most favorable adsorption site found for the single DBAQ on stepped Au(322) surface, the second DBAQ molecule is added, in a similar fashion as suggested for the flat surface, so as to form a 2D network.~As presented in Table \ref{T:celldim}, the terrace width of (322) surface is equal to 12.3 \AA, which is 2 \AA~larger than the supercell used for (111) surface.~It is important to mention here that this difference in the cell size affects both halogen- and hydrogen-bond strengths, due to the increased intermolecular distances. The optimized configuration of the 2D network together with the bond length scheme are shown in Fig. \ref{F:2dbaq+Au111}(c-d). As in the case of single-molecule adsorption, the flexibility of DBAQ facilitates the formation of continuous molecular carpeting on stepped surface. O$\cdot$$\cdot$$\cdot$H and Br$\cdot$$\cdot$$\cdot$H bond lengths vary from 2.96 to 3.68 \AA~and 3.03 to 3.46 \AA, as shown in Fig. S6. These bond distances are larger than those found on flat surface and of the sum of van der Waals radii for each element, indicating weaker hydrogen and halogen bonds formed on stepped (322) surface compared to flat (111) surface. Moreover, distances between Br and neighboring O atoms is too large ($\sim$4.71 \AA) to allow the formation of Br$\cdot$$\cdot$$\cdot$O bonds. Finally, we note that there is no significant difference found in adsorption heights between the monolayer and isolated molecule cases.

Compared to single DBAQ adsorbed on Au(322), the presence of a second molecule increases the adsorption energy by 0.16 eV/molecule, due to the Br$\cdot\cdot\cdot$H and O$\cdot\cdot\cdot$H bonds. Conversely, 0.1 eV energy is lost due to the different adsorption site of the second DBAQ molecule (see Fig. S3(c)) compared to the most favorable adsorption site. Hence, the flexibility of DBAQ, the higher reactivity of the step edge, and the formation of a network of Br$\cdot\cdot\cdot$H and O$\cdot\cdot\cdot$H bonds among adjacent molecules are the key players in the formation of stable continuous self-assembled monolayers on Au high-index surfaces.


\subsection{Electronic Properties of DBAQ on Au(111) and Au(322)} 

\begin{figure}[h!]
\centering
\includegraphics[width=75mm]{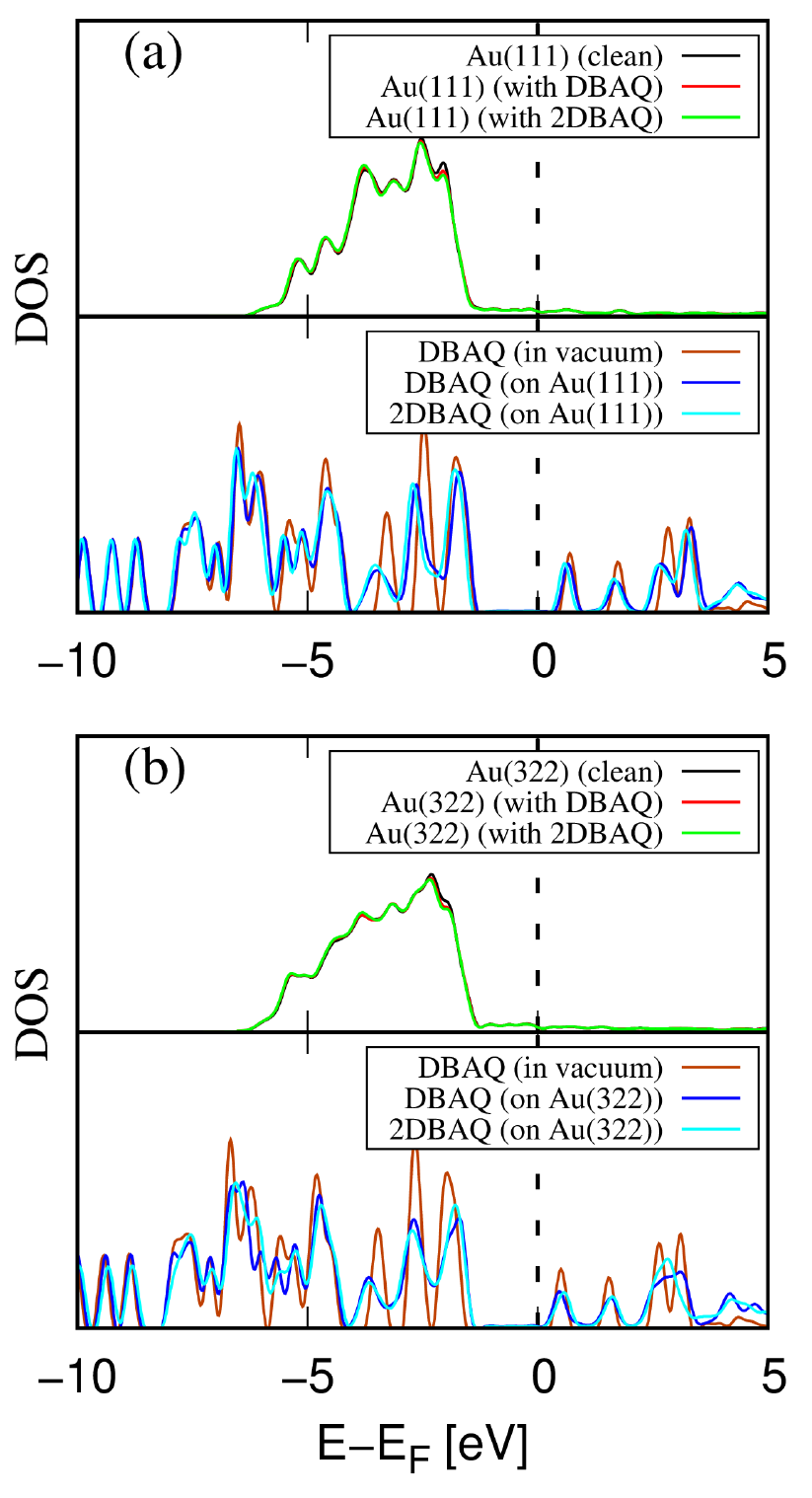}
\caption{PDOS of DBAQ and 2DBAQ on (a) Au(111) and (b) Au(322). The red and green lines in the upper panel show the surface gold atom PDOS when single and double DBAQ is adsorbed, and the black line indicates the PDOS of clean surface.~The blue and cyan lines in the lower panel display the PDOS of adsorbed single and double DBAQ on gold surfaces, respectively, and the red line indicates the PDOS of DBAQ in vacuum. 
The density of states are aligned setting the Fermi level to 0 eV. Energies are broadened by 0.2 eV.}
\label{F:PDOS}
\end{figure}

The electronic properties of the substrate highly depend on the surface type, which affects the binding characteristics of molecules and the formation of supramolecular structures. 
We have performed charge density and partial density of states (PDOS) analysis of the most favorable adsorption configurations, so as to further understand the interaction of DBAQ with Au surfaces.~PDOS is obtained by projecting the Kohn-Sham states on atomic orbitals, so that the contribution to the total electron density can be resolved in space and in orbital type. Fig. \ref{F:PDOS} shows the PDOS for single and two DBAQ molecules on Au(111) and Au(322) surfaces. The adsorption of DBAQ molecule(s) on flat and stepped gold surfaces does not significantly affect the electronic density of states of gold, in which the {\sl d}-states of the Au surface atoms prevail. In turn the electronic states of DBAQ are broadened and shifted by the interaction with gold. Additional molecule-molecule interactions at full coverage exhibit similar PDOS compared to single adsorbed cases. 

The contribution of the $p$-orbitals of individual species of atoms in DBAQ, in gas phase as well as adsorbed on Au, is displayed in Fig. S7. The highest occupied molecular orbital (HOMO) of DBAQ has $p$ character and entails contributions from carbon, bromine and oxygen.~When DBAQ is adsorbed on Au, the Br and C $p$-type electrons broaden and shift to lower energy. Hence their contribution to the frontier orbitals is reduced with respect to gas phase. The higher reactivity of the stepped surfaces leads to a stronger hybridization between the {\sl p}-electrons of the molecule and the {\sl s}-electrons of Au surfaces. The {\sl p}-states of C and O atoms below the Fermi level are significantly broadened and decreased in intensity, while the {\sl p}-states of Br lowered in intensity without a significant broadening.

\begin{figure}[h]
\centering
\includegraphics[width=85mm]{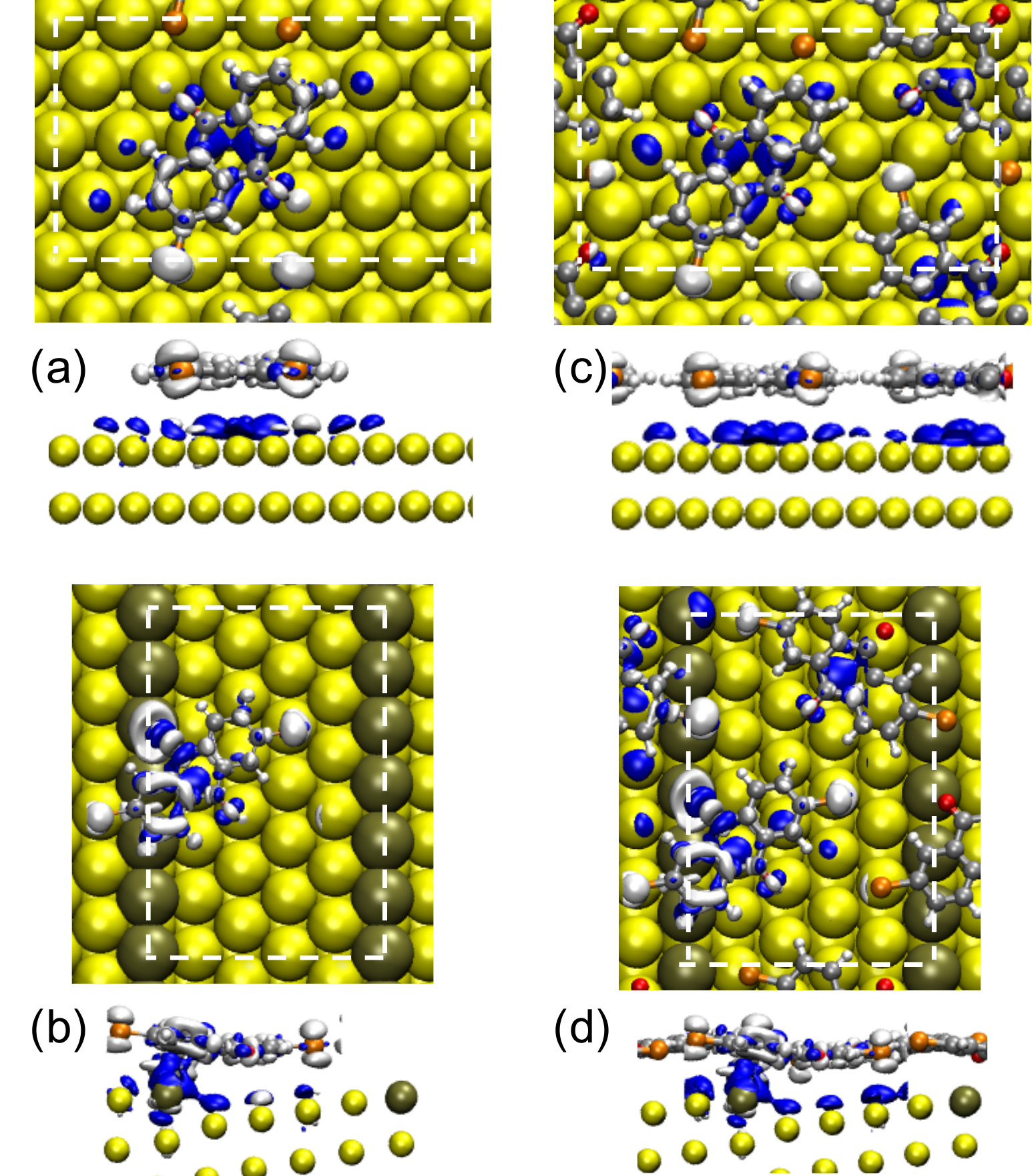}
\caption{3D isosurface of the electronic density differences with an isovalue of $\pm$0.003 e/\AA$^3$ of single DBAQ on (a) Au(111) with 6$\times$2 supercell and (b) Au(322), and two DBAQ on (c) Au(111) and (d) Au(322). Blue and white regions represent electron accumulation and depletion, respectively.}
\label{F:CDD}
\end{figure}

Fig. \ref{F:CDD} shows the electronic charge density difference for the energetically most favored structures on Au surfaces. The charge density differences due to the adsorption of DBAQ on flat and stepped surfaces are calculated as
\begin{equation}
\Delta \rho= \rhotot - \rhosur - \rhomol
\end{equation} 
where $\rhotot$ is the total charge density of DBAQ adsorbed on Au surface, $\rhosur$ is the charge density of Au surface in the adsorbed configuration without DBAQ molecule, and $\rhomol$ is the charge density of the isolated DBAQ in the adsorbed configuration. The average density profile, $\Delta$$\rho$$_{av}$(z), is also calculated for a quantitative analysis of the charge distribution and is presented in Fig. S8. When DBAQ is adsorbed on Au(111), the region between the surface and the molecule has charge accumulation (i.e., electron depletion, $\Delta$$\rho$$_{av}$(z) $<$ 0) and charge depletion (i.e., electron accumulation, $\Delta$$\rho$$_{av}$(z) $>$ 0) regions. Particularly the electron density is depleted just above the surface and there are accumulation regions below and above the molecule. Similar trends are found for the other configurations discussed in this work, which show that there is a notable interaction between the molecules and the surfaces. In general, Br atoms might have both positive and negative electrostatic potential regions when Br$\cdot$$\cdot$$\cdot$O and Br$\cdot$$\cdot$$\cdot$H bonds form, respectively \cite{Yoon2011}. For the systems studied here, Br atoms are the main electron depletion regions within the molecule due to the stronger interaction of Br atoms with hydrogens. 

The charge displacement ($\Delta Q = \int |\Delta \rho(z)| dz$) due to the adsorption of the molecule on the surface is quantitatively analyzed by integrating the charge density profile and summarized in Table \ref{T:dbaq+Au}.~As expected, stronger adsorption energies obtained for monolayer cases, compared to single molecule adsorption, yield larger $\Delta Q$ values.~Since the stepped surfaces are more reactive than flat ones, charge displacements calculated for flat surface are smaller by 0.28 and 0.39 $e$ for single and two DBAQ adsorbed configurations, respectively. \\

\section{Conclusions}
The interaction of 2,6-dibromoanthraquinones (DBAQ) molecule, which has been experimentally shown to induce a 2D supramolecular network on Au(111) surface, has been explored both on the flat (111) and stepped (322) surfaces of gold by means of density functional theory calculations, including the dispersion interactions in terms of the nonlocal vdW-DF correlation functional. Our calculations show that adsorptions at both surface steps and terraces are dictated by dispersion forces. The systematic search of various adsorption sites on surfaces reveals that, in spite of a slight preference for the B-30 site, the energy differences among different sites are small and DBAQ would diffuse easily at the surface.  
On flat surfaces DBAQ  retains its planar gas phase geometry, whereas when it adsorbs at step edges it is deformed. Such deformation energy compensates the energy gained by DBAQ interacting with highly reactive steps, thus making step and terrace adsorption energies almost equivalent. At monolayer coverage the intermolecular Br$\cdot\cdot\cdot$O and Br$\cdot\cdot\cdot$H halogen bonds and O$\cdot\cdot\cdot$H hydrogen bonds contribute to stabilize  DBAQ assemblies on flat and stepped surfaces, and lead to an adsorption energy increase by 0.21 eV/molecule with respect to isolated molecules. In contrast, the increased equilibrium distances of monolayer networks suggest weaker interaction of molecules with the surfaces, thus highlighting the interplay between these two types of interactions.

In general, this study demonstrates that the effect of surface type and the importance of van der Waals interactions as well as the role of halogen- and hydrogen-bonds cooperate to form two-dimensional self-assemblies  of complex aromatic molecules, such as anthraquinones on gold surfaces. Such interactions dictate the site and orientation selectivity of the adsorption of DBAQ molecule on gold, and, in general, determine the formation of supramolecular structures. Specifically the interplay among intermolecular bonding, molecular flexibility and surface adsorption foster the formation of continuous molecular carpets even at stepped surfaces, as observed in experiments.

\section*{Acknowledgment}
We are thankful to Se-Jong Kahng for useful discussions and for sharing experimental insight about this system. 
We acknowledge the provision of computational facilities and support by Rechenzentrum Garching of the Max Planck Society and access to the supercomputer JUQUEEN at the J\"{u}lich Supercomputing Center under project HMZ$33$. RP acknowledges TUBITAK ULAKBIM, High Performance and Grid Computing Center (TRUBA resources) for performing the numerical calculations reported.

\section*{Supplementary materials}
Adsorption geometries and energies of DBAQ molecule with different sites on Au(111) and Au(322) surfaces. Bond lengths of DBAQ molecules on surfaces. Projected density of states of atoms in vacuum and on gold surfaces. Average electronic density difference of systems.

\bibliography{dbaq_reference}

\end{document}